%% file: paper.tex
\documentclass[copyright]{eptcs}
\providecommand{\event}{PLACES'12} % Name of the event you are submitting to
\usepackage{breakurl}             % Not needed if you use pdflatex only.

\input{newmacros}

\input{localmacros}

\begin{document}

\title{Variant-Frequency Semantics for Green Futures}
\author{Yu David Liu \\
\institute{SUNY Binghamton \\ Binghamton, NY 13902, USA}
\email{davidL@cs.binghamton.edu}
}

\def\titlerunning{Variant-Frequency Semantics for Green Futures}
\def\authorrunning{Yu David Liu}

\maketitle

\lstset{classoffset=1, 
morekeywords={async},keywordstyle=\color{black}\bfseries, 
classoffset=0}

\begin{abstract}

This paper describes an operational semantics for futures, with the primary target on energy efficiency. The work in progress is built around an insight that different threads can coordinate by running at different ``paces,'' so that the time for synchronization and the resulting wasteful energy consumption can be reduced. We exploit several inherent characteristics of futures to determine how the paces of involving threads can be coordinated. The semantics is inspired by recent advances in computer architectures, where the frequencies of CPU cores can be adjusted dynamically. The work is a first-step toward a direction where variant frequencies are directly modeled as an essential semantic feature in concurrent programming languages.

\end{abstract}

\section{Introduction}

For software developers, adopting multi-core architectures is widely known to be a trade-off. On the benefit side, a programmable task -- if written as a multi-threaded program and deployed on multi-core platforms -- may potentially yield higher performance compared with a single-threaded implementation.  On the cost side however, correct and efficient multi-threaded programming is a complex matter. The vast majority of today's research on multi-core programming and compilation can be viewed as efforts to tip this cost-benefit analysis favorably, improving \emph{quality} of multi-core software:
\begin{center}
$\textrm{multi-core software quality} = \displaystyle \frac{\textrm{performance}}{\textrm{pain and horror of software development and use}}$
\end{center}
Examples include designing new programming models to ease programming efforts and enforce invariants, new program analyses to find concurrency bugs, or new optimization techniques to further improve performance.
%\paragraph{Energy Consumption as Costs}

\paragraph{Multi-Core Software Energy Efficiency}

An additional form of cost -- obvious but so far largely under the radar of multi-core programming and compilation research -- is the energy consumption of multi-core architectures:
 the energy consumption of CPUs multiplies when platforms evolve from single-core architectures to multi-core ones due to the inherent nature of digital circuits. 
In this paper, we call for more research efforts to tip a new flavor of cost-benefit analysis favorably, improving \emph{energy efficiency} of multi-core software:

\begin{center}
$\textrm{multi-core software energy efficiency} = \displaystyle \frac{\textrm{performance}}{\textrm{energy consumption}}$
\end{center}

To come up with a software-centered solution for energy efficiency, it is important to identify energy inefficiencies as \emph{introduced by software}. Indeed, when we deploy a multi-threaded program on a 20-core machine, we would have tolerated a 20x increase of energy consumption if our program yields 20x speed-up. The culprit that prevents this -- according to the now famous Amdahl's law \cite{amdahl} -- is really the program (algorithm) itself! The law tells us linear speed-up is impossible on multi-core executions for algorithms with any serial components (which by the way, apply to virtually all practical programs). Performance degrades the most when a parallel execution is stalled due to the need for executing the program's serial components. To improve energy efficiency, it is thus the most profitable if we focus on minimizing energy consumption for these parallelism-stalling oprations. 

On the programming language level, such operations are often realized through synchronization primitives. When two threads synchronize, the first thread arriving at the synchronization point needs to wait for the arrival of the second. Operationally, the intuitive notion of ``wait'' translates to either spinning or blocking \cite{ArtOfMpP} of the first thread. 
Unfortunately, neither spinning nor blocking is energy-efficient. Spinning -- also known as busy waiting -- consumes energy with no execution throughput directly related to program code. Blocking -- the strategy that context-switches the first thread so that the CPU core can be occupied by other threads -- increases overall CPU utilization but comes with the cost of context switch. This especially takes a toll on energy consumption: context switch usually leads to significant reduction on cache locality; the resulting cache misses are known to be one of the most expensive operations in terms of energy consumption.

\paragraph{Energy-Efficient Futures}

This paper puts the spotlight on one particular form of synchronization mechanism, futures \cite{multilisp}, and argues that several of their distinct traits -- if exploited -- can potentially improve energy efficiency of multi-core program execution. Our key insight is that, to avoid the useless energy consumption of spinning or blocking, different threads can execute at different ``paces,'' so that the thread likely to arrive early ``saunters'' to the synchronization point whereas the one likely to arrive late ``sprints'' to the synchronization point. To achieve the effect of sauntering and sprinting is not hard: modern CPUs are almost invariably equipped with abilities to dynamically adjust frequencies and voltages, a strategy widely known as Dynamic Voltage and Frequency Scaling (DVFS) \cite{dvs}. The main challenge here is to \emph{determine which thread should saunter, and which thread should sprint}, a question that will be answered in the next section.

\section{Green Futures: The General Approach}

Futures have long known to be appealing for implicit thread management \cite{multilisp} and program optimization \cite{Flanagan:future}. The idea was popularized in a functional setting (such as MultiLisp and Scheme), and later successfully adopted to object-oriented languages, both as research prototypes \cite{hicksfuture,safefuture} and mainstream language extensions (Java and C\#). For example, the following pseudo-code demonstrates the use of futures in a Java-like method: 

\lstset{language=java, keywordstyle=\color{black}\bfseries, numbers=left, numberstyle=\tiny, stepnumber=1, numbersep=5pt,stringstyle=\ttfamily} 

%\lstset{emph={[2]SOME_TO_EMPHASIZE}, emphstyle={[2]\color{red}}}

\lstset{classoffset=1, 
morekeywords={future},keywordstyle=\color{black}\bfseries, 
classoffset=0}

\lstset{moredelim=*[is][\color{red}]{(((}{)))}}

\lstset{escapeinside={(*@}{@*)}}

\begin{center}
\begin{tabular}{c}
\begin{lstlisting}
void procRequest(Socket s) {
  Buffer in = future readBuf(s); (*@\label{l:future-1}@*)
  ... (*@\label{l:future-3}@*)
  int size = in.position(); (*@\label{l:future-2}@*)
  ...
} 
\end{lstlisting}
\end{tabular} 
\end{center}

Here, keyword $\textbf{future}$ signifies that the invocation to $\texttt{readBuf}$ at L.~\ref{l:future-1} is an asynchronous thread creation (called a \emph{future creation}); the method body of $\texttt{readBuf}$ will be executed in a separate \emph{future thread}, and the program counter in the thread executing $\texttt{procRequest}$ (the \emph{parent thread}) immediately moves on to the next statement. From this moment on, the two threads will run in parallel, with the future thread serving as a ``producer'' which eventually fulfills the return value of $\texttt{readBuf}$ -- called \emph{future realization} (or \emph{fulfillment}) -- and the parent thread serving as a ``consumer'' when the return value of $\texttt{readBuf}$ is needed -- $\texttt{in.position()}$ invocation here at  L.~\ref{l:future-2} -- called \emph{future claim} (or \emph{touch}). 
With the method body of $\texttt{readBuf}$ and the omitted statements at L.~\ref{l:future-3} running in parallel, futures offer an appealingly simple and incremental way to speed up previously serial code (\emph{i.e.} the one when the keyword $\textbf{future}$ is removed).

To improve energy efficiency, we design a variant-frequency execution strategy for the two threads involved in futures. Specifically, we adopt the following general strategy: 

\begin{quote}
\textbf{Main Strategy}: we set the parent thread executing at a \emph{lower} frequency level than the future thread. 
\end{quote}

This strategy is designed thanks to two distinctive traits of futures. First, futures shapes up a fundamentally asymmetric relationship between two threads: one is a producer, and the other is a consumer; Second, the future thread terminates upon future realization. Let us now demonstrate why the \textbf{Main Strategy} is a sensible one. Not to lose generality, observe that there can only be two cases for a future-involved execution:

 \begin{center}
 \parbox[c]{.8\columnwidth}{\includegraphics[width = .8\columnwidth]{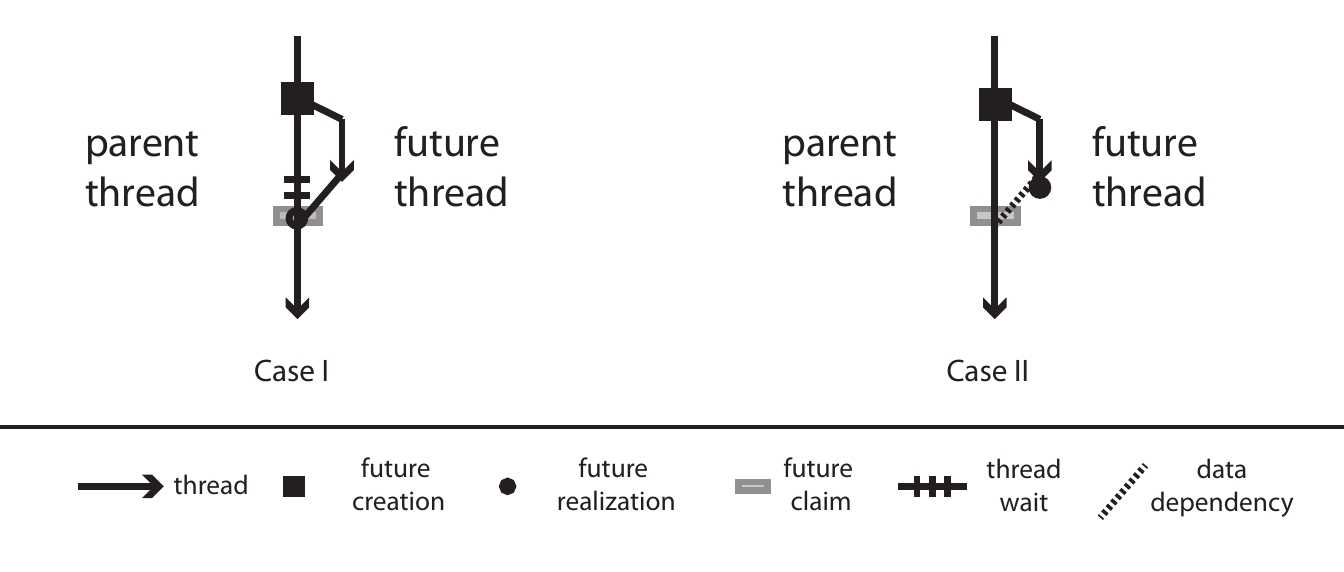}} 
\end{center}

In Case I, despite having the parent thread executing at a lower frequency, it still reaches the future claim point before the future is realized. In this case, the parent thread does need to spin or block, but observe that the duration of spinning or blocking -- hence useless energy consumption -- is reduced compared with the scenario where the parent had chosen to run faster and reached the claim point even earlier. In Case II, as a result of the ``faster'' execution, the future thread successfully fulfills the future before the parent thread claims it. In this case, the future thread has accomplished its mission of being, and can be terminated. No spinning or blocking is needed. When the parent thread finally reaches the claim point, the value (such as \texttt{in} in the example) is ready. No spinning or blocking is needed for the parent thread either.

It should be noted that what really matters is the \emph{relative} pace of the parent thread and the future thread, not the absolute one. For instance, instead of slowing down the parent thread, the reasoning above still stands if an implementation chooses to \emph{speed up} the future thread, or slowing down the parent thread and speeding up the future thread at the same time. All variations are likely to improve on overall energy efficiency -- in that the likelihood and duration of wasteful wait is reduced -- but they might have different effects on performance and overall energy consumption. For instance, if one chooses to (absolutely) slow down the parent thread, Case II above has the potential to lead to performance penalties because the program may run longer as a result of the slower execution of the critical path (the parent thread). On the other hand, if one chooses to (absolutely) speed up the future thread, Case II will not have the aforementioned negative impact on performance. Overall, what this suggests is the same \textbf{Main Strategy} above may lead to different implementation choices in DVFS. In Sec.~\ref{sec:formal}, we provide a more precise account of this approach. % In Sec.~\ref{sec:experiment}, we discuss a number of experimental issues.

According to our preliminary experiments, the overhead of DVFS -- usually within tens of micro-seconds in existing architectures -- can be ignored. Indeed, threads usually execute at a duration magnitudes longer, otherwise the cost of thread management would have invalidated their \emph{raison d'\^ete}.

\section{Operational Semantics}
\label{sec:formal}

In this short presentation, we illustrate our ideas through a mini-language, a multi-threaded addition calculator: 

\figurebox{
\begin{math}
\begin{array}{l l l r}

e & \defassign & e + e \mid \kwfuture\ e  \mid v & \textit{expressions} \\

v & \defassign & \iconst \mid \fv & \textit{values} \\

\parconfig & \defassign & \fuclosure{\freq}{e}{\fv} \parallel \parconfig & \textit{configurations} \\

\end{array} 
\end{math}
}
Values are either integers or future values ($\fv$). A new thread can be created via the $\kwfuture\ e$ expression, and future claim may happen for the $+$ expression when either of its arguments is a a future value. A parallel configuration is formed by concatenating single-threaded computations together, via commutative $\parallel$. Each single-threaded computation takes the form of $\fuclosure{\freq}{e}{\fv}$, denoting expression $e$ is currently evaluated on a CPU with frequency $\freq$, for realizing a future value $\fv$. In addition, let us define the frequencies supported by each CPU core as a finite well-ordered set $W = \{\freq_1, \dots, \freq_n\}$ where $\freq_1 < \freq_2 \dots < \freq_n$ (as in hertz). Since this set is fixed given a hardware environment, the rest of the definitions are implicitly parameterized by $W$.

The small-step operational semantics is defined by the transitive reduction relation $\cloreduct{}$ over configurations. The reduction rules are defined as follows:

\figurebox{

 \begin{tabular*}{\textwidth}{l} 
$
\begin{array}{rr@{\ }c@{\ }l@{\ }lll} 
%%%%%%%%%%%%%%%%%%%%%%%
\rtitle{Create} & \fuclosure{\freq}{ \env [ \textbf{future}\ e]}{ \fv'} 
& \cloreduct{} &
\fuclosure{\upf \freq}{e}{ \fv} \ \parallel \ \fuclosure{\downf \freq}{ \env[\fv]}{ \fv'}   &  \textrm{ if } &  \textrm{$\fv$ fresh} \\
\rtitle{Claim} & \fuclosure{\freq}{\claimenv[\fv]}{\fv'} \parallel \ \fuclosure{\freq'}{v}{\fv}  &\cloreduct{} \ &  \fuclosure{\claimf \freq}{\claimenv[v]}{ \fv'}  \\
\rtitle{Add} & \fuclosure{\freq}{ \env [ i + i']}{ \fv} 
& \cloreduct{} & \fuclosure{\freq}{ \env [ i'']}{ \fv} & \textrm{ if } & i'' \textrm{ sum of } i, i' \\
\rtitle{Cxt} & c \parallel c'' & \cloreduct{} & c' \parallel c'' &  \textrm{ if } & c \cloreduct{} c' \\[3ex]
\end{array} 
$
\end{tabular*}

\begin{math}
\begin{array}{l l l r}

\env & \defassign & \bullet \mid \env + e \mid v + \env & \textit{evaluation context} \\
\claimenv & \defassign & \bullet \mid \claimenv + v \mid i + \claimenv & \textit{claim context} \\

\end{array} 
\end{math}

}

The main novelty here is that the frequency of each thread execution can be explicitly adjusted, through unary operators for upscaling ($\upf$ and $\claimf$) and downscaling ($\downf$). How these operators are defined concretely is the standard problem of scaling factor selection in DVFS. Some design choices, including the difference between $\upf$ and $\claimf$, will be discussed shortly. \rtitle{Create} and \rtitle{Claim} correspond to future creation and future claim, respectively. At future creation time, the frequency for the ``future'' thread is scaled up, whereas the `parent'' thread is scaled down. This design choice reflects our main principle of frequency adjustment: the ``future'' thread should hurry up to realize the future values, whereas the ``parent'' thread should leisurely proceed. \rtitle{Claim} shows that future claims are fundamentally an operation of synchronization. Here, the blocking semantics is used, whenever a future value needs to be claimed, the reduction cannot progress until the future is realized. After future claim, the frequency of the ``parent'' thread needs to scale up -- intuitively, the reason for the ``parent'' thread to saunter no longer exists.  In addition to the standard evaluation context $\env$, a separate claim context $\claimenv$ is defined, for execution configurations where a future must be realized. Note that the definition here is able to support ``futures of futures'': it is possible that a future thread realizes its future with another future value -- in which case the $v$ metavariable in \rtitle{Claim} is a future value in its own. 

For example, if a (somewhat contrived) program $3 + \textbf{future}\ \textbf{future}\ (3+4)$ starts its execution at frequency $\freq_{\tt init}$, the following reduction sequence is possible, where $\fv_{\tt init}$ is a trivial future value for the initial configuration, and  $\fv_1$, $\fv_2$ are fresh:

%\begin{center}
\hskip -5ex \begin{tabular*}{\textwidth}{l} 
$
\begin{array}{llr} 
%%%%%%%%%%%%%%%%%%%%%%%
& \fuclosure{\freq_{\tt init}}{ 2 + \textbf{future}\ (\textbf{future}\ (3+4))}{ \fv_{\tt init}}  & \rtitle{Create}\\
\cloreduct{} & \fuclosure{\upf \freq_{\tt init}}{\textbf{future}\ (3+4)}{ \fv_1} \ \parallel \ \fuclosure{\downf \freq_{\tt init}}{ 2+\fv_1}{ \fv_{\tt init}}   & \rtitle{Create}, \rtitle{Cxt} \\
\cloreduct{} & \fuclosure{ \upf (\upf \freq_{\tt init})}{3+4}{ \fv_2} \ \parallel \fuclosure{\downf (\upf \freq_{\tt init})}{\fv_2}{ \fv_1} \ \parallel \ \fuclosure{\downf \freq_{\tt init}}{ 2+\fv_1}{ \fv_{\tt init}}  & \rtitle{Add}, \rtitle{Cxt}   \\
\cloreduct{} & \fuclosure{ \upf (\upf \freq_{\tt init})}{7}{ \fv_2} \ \parallel \fuclosure{\downf (\upf \freq_{\tt init})}{\fv_2}{ \fv_1} \ \parallel \ \fuclosure{\downf \freq_{\tt init}}{ 2+\fv_1}{ \fv_{\tt init}}    & \rtitle{Claim}, \rtitle{Cxt} \\
\cloreduct{} &  \fuclosure{\claimf(\downf (\upf \freq_{\tt init}))}{7}{ \fv_1} \ \parallel \ \fuclosure{\downf \freq_{\tt init}}{ 2+\fv_1}{ \fv_{\tt init}}    & \rtitle{Claim}, \rtitle{Cxt} \\
\cloreduct{} &  \fuclosure{\claimf(\downf \freq_{\tt init})}{ 2+7}{ \fv_{\tt init}}    & \rtitle{Add} \\
\cloreduct{} &  \fuclosure{\claimf(\downf \freq_{\tt init})}{9}{ \fv_{\tt init}}    & \\
\end{array} 
$
\end{tabular*}
%\end{center}

\paragraph{Scaling Factor Selection}
\label{sec:scale}

One possible way of defining the upscaling/downscaling operators is to apply the standard functions of computing successive and preceding elements over $W =  [\freq_1, \dots, \freq_n]$: 

\begin{center}
$
\begin{array}{rcll}
 \upf \freq_k \df \claimf \freq_k & \df &  \freq_{k+1} &  1 \leq k \leq n-1 \\
 \upf \freq_n \df \claimf \freq_n  & \df &  \freq_n & \\
\downf \freq_k & \df &  \freq_{k-1} &  2 \leq k \leq n \\
\downf \freq_1 & \df &  \freq_1 & \\

\end{array}
$
\end{center}

Here, the future thread is literally scaled up and the parent is scaled down. If a parent thread is going to create two future threads, the second future thread is going to execute at a lower frequency than the first (because the parent thread itself has been scaled down after the first future creation). Assuming all futures created by the parent thread will be claimed, the parent thread eventually will return to the original frequency. As another strategy, we can adjust the frequency of the parent thread only:

\begin{center}
$
\begin{array}{rcll}
\upf  \freq_k & \df &  \freq_k &  1 \leq k \leq n \\
\claimf  \freq_k & \df &  \freq_{k+1} &  1 \leq k \leq n-1 \\
\claimf \freq_n & \df & \freq_n \\
\downf \freq_k & \df &  \freq_{k-1} &  2 \leq k \leq n \\
\downf \freq_1 & \df & \freq_1 
\end{array}
$
\end{center}

\section{Future Work}
\label{sec:experiment}

This paper describes a work in progress, focusing on the ideas. A full-fledged semantics is under development. In particular, it remains to be seen how future safety \cite{safefuture} interacts with the proposed ideas in an imperative setting. The fact that multiple scaling factor selection strategies exist clearly demonstrates the importance of experimental methods in this project. For each selection strategy, we are interested in exploring its impact on both performance -- including both the spinning/blocking time and the overall execution time of the program -- and energy consumption, and measuring it in a more rigorous setting \emph{e.g.} through Energy-Delay Product \cite{gonzalez_horowitz} and Energy-Delay Squared Product \cite{Martin:2002:EMT:783060.783076}.

This paper demonstrates that a compiler decision on DVFS can be made to improve the energy efficiency of multi-threaded programs without the knowledge of their logical/execution details. Like most optimization problems, the more knowledge one has on the optimization space, the more effective/optimal the solution will be. An interesting direction is to see how the general principle discussed in this paper can be further combined with static/dynamic information of programs to contribute to additional energy efficiency. For example, instead of viewing the described algorithm here as one where all scaling points and scaling factors are entirely fixed at compile time and oblivious of the run-time behaviors, an adaptive algorithm can be designed where the concrete definitions of $\upf$, $\claimf$, and $\downf$ can be adjusted at run time based on run-time environment information and program profiling data.

\paragraph{Acknowledgment}

We thank the anonymous reviewers for their useful suggestions.
This work is being supported by NSF CAREER Award CCF-1054515 and a Google Faculty Award. 

% use comma to separate if more.
\bibliography{../../../mylib/bibs/multithread,../../../mylib/bibs/energy}

\end{document}

%% file: newmacros.tex
\usepackage{times}
\usepackage{amsfonts}
\usepackage{amsmath} %xrightarrow
\usepackage{amssymb} %leadsto rightsquigarrow rhd mathbb Join Subset
\usepackage{array}
\usepackage{color}
\usepackage{pifont}
\usepackage{epsfig}
\usepackage{alltt}
\usepackage{latexsym}
\usepackage{slashbox}
\usepackage{mathpartir}
\usepackage{wasysym} %rhd, RHD
\usepackage{textcomp}
\usepackage{stackrel} % adds stack symbol below arrow option
\usepackage{setspace}
\usepackage{stmaryrd}
\usepackage{amsbsy} % bold greek letter
\usepackage{mathtools} % newtagform
\usepackage{ifthen}
\usepackage{boxedminipage} 
\usepackage{bbold}
\usepackage{turnstile}
\usepackage{balance,wrapfig,listings,caption,subcaption}
\usepackage{url,hyperref, graphicx,listings}

\def\df{\ \stackrel{\rm def}{=}\ }

\def\defassign{\!\!::=\!\!}

\def\cloreduct#1{\stackrel{#1}{\Longrightarrow}}

\def\env{\mathbf{E}}

\def\partial{\mathcal{LU}}

\def\augG#1#2#{ #1 \lhd  #2}

\def\rtitle#1{{\sf (\textrm{R-#1})}}

\newcommand{\figurebox}[1]{
\begin{center}
\noindent\begin{minipage}[c]{\columnwidth} 
\begin{center}
\vskip 0ex
\begin{tabular*}{\columnwidth}{c}\end{tabular*}
\begin{tabular*}{\columnwidth}{c}\hline\end{tabular*}
\vskip 1ex
#1
\vskip 1ex
\begin{tabular*}{\columnwidth}{c}\hline\end{tabular*}
\vskip 2ex
\end{center}
\end{minipage} 
\end{center}

}

\definecolor{mygray}{cmyk}{0.1,0.3,0.1,0.50}

\long\def\omitthis#1{}

\long\def\inproof#1{}

%% file: localmacros.tex
\newcommand{\kwfuture}{\texttt{\bf future}}

\newcommand{\freq}{\textsf{fq}}
\newcommand{\upf}{\uparrow}
\newcommand{\claimf}{\Uparrow}
\newcommand{\downf}{\downarrow}

\newcommand{\claimenv}{\mathbf{C}}

\newcommand{\fv}{\emph{fv}}
\newcommand{\iconst}{\emph{i}}

\newcommand{\parconfig}{\emph{c}}

\newcommand{\fuclosure}[3]{\texttt{\bf cl}(#1,#2)^{#3}}